\documentclass[aps,pre,reprint,groupedaddress]{revtex4-1}

\usepackage{graphicx}
\usepackage{amsmath}

\begin{document}


\renewcommand{\d}{\partial}

\title{Diffusion Enhancement of Brownian Motors  
Revealed by a Solvable Model}


\author{Ryo Kanada}
\affiliation{%
Department of Biophysics, 
Graduate School of Science, Kyoto University, Kyoto 606--8502, Japan}
\affiliation{%
Compass to Healthy Life Research Complex Program, RIKEN, 
6-5-3 Minatojima-Minamimachi, Chuo-ku, 
Kobe 650--0047, Japan.}

\author{Ryota Shinagawa}
\author{Kazuo Sasaki}
\email{sasaki@camp.apph.tohoku.ac.jp}
\affiliation{%
Department of Applied Physics, Graduate School of Engineering, 
Tohoku University, Senda 980--8579, Japan
}%


\date{\today}

\begin{abstract}
A solvable model is proposed and analyzed to reveal the mechanism underlying 
the diffusion enhancement recently reported for a model of molecular motors and 
predicted to be observed in the biological rotary motor $\rm F_1$-ATPase.  
It turns out that the diffusion enhancement for the present model can 
approximately described by a random walk in which the waiting time for a step to 
occur is exponentially distributed and it takes nonzero time to proceed forward by the step. 
It is shown that the diffusion coefficient of such a random walk can significantly 
be increased when the average waiting time is comparable to the average stepping time. 
\end{abstract}


\maketitle


\section{Introduction}
\label{sec:intro}

There can be various ways to produce an effective diffusion coefficient 
larger than what is expected from the Einstein relation. 
A classical example of such \textit{diffusion enhancement} is the swimming of bacteria~\cite{berg93}. 
A bacterium swims with a constant speed and occasionally changes its swimming direction.  
The resulting motion is a random walk with an effective diffusion coefficient 
much larger than that of the Brownian motion it would undergo if it stopped swimming. 
This diffusive motion helps bacteria find their foods or move away 
from harmful environments, for example. 
Recently, it was found that a Brownian particle moving in a one-dimensional periodic potential 
exhibits the diffusion enhancement under a constant external force of magnitude 
close to the maximum slope of the potential~\cite{costantini99, reimann01, reimann02}. 
This phenomenon was observed experimentally in a colloidal system~\cite{speer12},  
a biomolecule having a rotating subunit~\cite{hayashi15}, 
and DNA diffusing in an array of entropic barriers~\cite{kim17}. 
The diffusion enhancement also occurs in on-off ratchets~\cite{germs13}  
in which an asymmetric, periodic potential for colloidal particles is switched on and 
off periodically, if the duration of potential-off interval is such that the root-mean square displacement 
of the particle by free diffusion in this interval is comparable to the periodicity of the potential. 

In our previous work~\cite{shinagawa16}, 
it is demonstrated theoretically that the diffusion enhancement can occur in molecular motors 
that move autonomously by consuming free energy available from the chemical reaction 
catalyzed by themselves if a constant external force of appropriate magnitude is applied. 
In particular, it is suggested that the diffusion enhancement can be observed in the F$_1$-ATPase, 
a biological rotary motor, which catalyzes the hydrolysis of adenosine triphosphate (ATP).  
It has turned out that the mechanism of enhancement in the case of high ATP concentration 
is essentially the same as the one for the particle in a tilted periodic 
potential~\cite{costantini99, reimann01, reimann02}. 
On the other hand, the mechanism in the case of low ATP concentration has not been clarified yet. 
The purpose of the present work is to study the diffusion of a simplified model of molecular motors 
to elucidate the mechanism of diffusion enhancement characteristic of chemically driven systems. 
In what follows effective diffusions coefficient will simply be called diffusion coefficients. 

The models considered here and in the previous work~\cite{shinagawa16} 
are of ratchet type~\cite{julicher97, reimann02pr, kawaguchi14}, 
in which a moving part of the motor (e.g., the rotor in a rotary motor) 
is represented by a Brownian particle subject to a potential, 
which is switched to another upon a chemical transition associated with 
the reaction catalyzed by the motor. 
In the model used in the previous paper~\cite{shinagawa16}, 
an external force, as well as rate constants, can control the transition rates 
because the transition rates are assumed to depend on the particle position, 
which is affected by the force; 
the dependence of the diffusion coefficient on the force for given rate constants 
exhibits enhancement in a certain range of the force. 
By contrast, an external force is not included in the model of the present paper, 
and a rate constant is varied to study the diffusion enhancement. 
Another simplification is that chemical transitions are supposed to take place 
only when the particle is located at particular points, 
which enable us to obtain a closed-form expression for the diffusion coefficient. 

The paper is organized as follows. 
The model is introduced in the next section, 
and the closed-form expressions for the velocity and diffusion coefficient 
are given in Sec.~\ref{sec:closed}. 
Explicit calculations of the diffusion coefficient are carried out for a model 
with piecewise linear potentials in Sec.~\ref{sec:plinear}, 
where the diffusion enhancement is demonstrated. 
In Sec.~\ref{sec:pwalk}, we discuss the mechanism of the diffusion enhancement 
observed in Sec.~\ref{sec:plinear} on the basis of a simple random walk, 
which we call an \textit{extended Poisson walk}. 
Concluding remarks will be given in Sec.~\ref{sec:conclusion}.  
Some of the details of calculations and expressions are given in Appendices. 

\section{Potential-switching ratchet}
\label{sec:retchet}

\begin{figure}[htb]
\includegraphics[scale = 0.95]{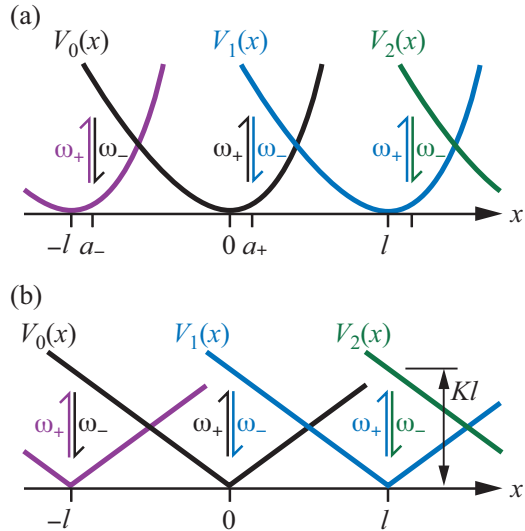}%
\caption{\label{fig:model}%
	(Color online) 
	Generic (a) and specific (b) models of molecular motors considered 
	in this work. 
	The motor is represented by a Brownian particle subject to one of the potentials 
	$V_n(x)$ at a time. The potential is switched from $V_n$ to $V_{n+1}$ or 
	$V_{n-1}$ stochastically with rates proportional to $\omega_{+}$ and $\omega_{-}$, 
	respectively, if the particle is located at $x = a_{+} + nl$ or $a_{-} + nl$; 
	we set $a_{+} = 0$ and $a_{-} = -l$ for~(b).  
}%
\end{figure}

We consider a variant of pulsating ratchets as a model of 
biological molecular motor~\cite{julicher97, reimann02pr, kawaguchi14}. 
The motor is modeled as a Brownian particle moving in one dimension, along the $x$ axis,  
subject to potentials $V_n(x)$ ($n = 0, \pm1, \pm2, \dots$) of identical shapes 
arranged periodically with period $l$ as shown in Fig.~\ref{fig:model}(a), 
i.e., they satisfy
\begin{equation}
\label{eq:vn}
	V_{n+1}(x) = V_n(x - l)
	\quad (n = 0, \pm1, \pm2, \dots).
\end{equation}
The potentials are assumed to be unbounded above. 
Only one of the potentials acts on the particle at a time, say $V_n$, 
and it is stochastically switched to $V_{n+1}$ or $V_{n-1}$. 
The motor will be said to be in \textit{state} $n$ if potential $V_n$ is acting.  
The dynamics of the particle is assumed to be over-damped. 
The diffusion coefficient of the particle in the absence of the potentials 
is given by the Einstein relation $D_0 = k_\text{B}T/\gamma$, 
where $\gamma$ is the coefficient of the drag force on the particle 
from the surrounding fluid, $T$ is the temperature, 
and $k_\text{B}$ is the Boltzmann constant. 

Let $P_n(x,t)\,dx$ be the probability to find the motor in state $n$ and 
the particle in the interval $(x, x + dx)$ on the $x$ axis at time $t$, 
and $w_n^\pm(x)$ be the rate of transition from state $n$ to state $n \pm 1$ 
when the particle is located at $x$.  
Then the time evolution of $P_n(x,t)$ is described by the Fokker-Planck equations, 
\begin{align}
\label{eq:fpeq}
	\frac{\d P_n}{\d t} +\frac{\d J_n}{\d x}
	&= -(w_n^{+} + w_n^{-})P_n 
	\nonumber\\
	&\quad{} + w_{n-1}^{+}P_{n-1} + w_{n+1}^{-}P_{n+1}, 
\end{align}
where $J_n$ is the probability current in state $n$ defined by 
\begin{equation}
\label{eq:jndef}
	J_n \equiv -D_0 \left(\frac{\d}{\d x} + \frac{dU_n}{dx}\right)P_n
\end{equation}
with  
\begin{equation}
\label{eq:undef}
	U_n(x) \equiv V_n(x)/k_\text{B}T 
\end{equation}
being the dimensionless potential. 
We remark that the arguments given in this section applies also to the case in which 
a constant external force $F$ is applied to the particle 
if the right-hand side of Eq.~(\ref{eq:undef}) is replaced by $[V_n(x) - Fx]/k_\text{B}T$. 

We assume that the transition from one state to another takes place  
when the particle is located at a particular position 
(this corresponds to the idea that the change in chemical state 
of a motor protein occurs when it is in a particular conformation~\cite{julicher97}), 
and adopt the following expressions for $w_n^{\pm}(x)$: 
\begin{equation}
\label{eq:wnpm}
	w_n^{\pm}(x) = \omega_\pm\delta(x - a_\pm - nl), 
\end{equation}
where $\delta(x)$ is the delta function, $\omega_\pm$ are positive constants, 
and $a_{+}$ and $a_{-}$ are constants satisfying $a_{-} = a_{+} - l$; 
see Fig.~\ref{fig:model}(a). 
The transition $n \to n+1$ and its reversal occur at $x = a_{+} + nl = a_{-} + (n+1)l$. 
Supposing that the ``forward'' transition $n \to n+1$ is triggered by 
a chemical reaction by which the free energy of the environment is decreased by 
$\Delta\mu$, we have the relation 
\begin{equation}
\label{eq:dbalance}
	\omega_{+}/\omega_{-} 
	= \exp\left[U_0(a_{+}) - U_0(a_{-}) + \Delta\mu/k_\text{B}T\right]
\end{equation}
from the condition of local detailed balance.

The velocity $v$ and the diffusion coefficient $D$ of the motor are defined by 
\begin{equation}
\label{eq:vddef}
	v \equiv \lim_{t \to \infty}\frac{\langle x(t) - x(0)\rangle}{t}, 
	D \equiv \lim_{t \to \infty}\frac{\langle [x(t) - x(0) - vt]^2\rangle}{2t}, 
	\quad
\end{equation}
where $x(t)$ is the location of the particle (motor) at time $t$ 
and the angular brackets indicate the statistical average. 
The velocity can be obtained from the  steady-state solutions $P_n(x)$ 
of the Fokker-Planck equations~(\ref{eq:fpeq}), 
which satisfy the ``periodicity condition'' $P_n(x) = P_0(x - nl)$.   
Let $P(x)$ be the rescaled $P_0(x)$ so that it satisfies 
\begin{equation}
\label{eq:pnorm}
	\int_{-\infty}^\infty P(x)\,dx = 1. 
\end{equation} 
Then we have~\cite{julicher97, reimann02pr}
\begin{equation}
\label{eq:vformula}
	v = \int_{-\infty}^\infty f(x)P(x)\,dx, 
	\quad
	f(x) \equiv -D_0\,dU_0(x)/dx. 
\end{equation}
To calculate the diffusion coefficient, we need to obtain the 
auxiliary function $Q(x)$ that satisfies 
\begin{align}
	&\frac{d}{dx}\left[D_0\frac{d}{dx} - f(x)\right]Q(x) - \left[w_0^{+}(x) + w_0^{-}(x)\right]Q(x)
	\nonumber \\
	&\qquad{}+ w_0^{+}(x + l)Q(x + l) + w_0^{-}(x - l)Q(x - l)
	\nonumber \\
	&\quad{}= \left[v - f(x) + 2D_0\frac{d}{dx}\right]P(x) 
\label{eq:qeq}
\end{align}
and the boundary condition that $Q(x) \to 0$ as $|x| \to \infty$. 
The diffusion coefficient is calculated as~\cite{harms97, sasaki03, shinagawa16}
\begin{equation}
\label{eq:dformula}
	D = D_0 + \int_{-\infty}^\infty \left[f(x) - v\right]Q(x)\,dx.
\end{equation}

\section{Closed-Form Expressions}
\label{sec:closed}

The specific functional forms of $w_n^\pm(x)$ given in Eq.~(\ref{eq:wnpm}) enable us to 
derive closed form expressions for $v$ and $D$, as explained in Appendix~\ref{sec:derivation}. 
To express the result for $v$ concisely, 
we introduce constants $\zeta$, $\phi_0$, $\phi_1$, and $u_\pm$ defined by
\begin{align}
	&\zeta \equiv \int_{-\infty}^\infty e^{-U(x)}\,dx, \quad
	\label{eq:zetadef}
	\\
	&\phi_0 \equiv \int_{a_{-}}^{a_{+}} e^{U(x)}\,dx,
	\label{eq:phi0def}
	\\
	&\phi_1 \equiv\int_{a_{-}}^{a_{+}} e^{U(y)}\,dy\int_{-\infty}^y e^{-U(x)}\,dx,
	\label{eq:phi1def}
	\\
	&u_\pm \equiv \omega_\pm\exp\left[-U(a_\pm)\right],   
	\label{eq:upmdef}
\end{align}
where 
\begin{equation}
\label{eq:uxdef}
	U(x) \equiv U_0(x). 
\end{equation}
Then, the velocity is expressed as 
\begin{equation}
\label{eq:vresult}
	v = \frac{(u_{+} - u_{-})D_0l}{\zeta(D_0 + \phi_0u_{-}) + \phi_1(u_{+} - u_{-})}.
\end{equation}
Note that we have $u_{+} > u_{-}$ for $\Delta\mu > 0$ 
according to the detailed-balance condition~(\ref{eq:dbalance}) 
and that the denominator in Eq.~(\ref{eq:vresult}) is positive 
since $\zeta\phi_0 - \phi_1 > 0$, which can be verified from Eqs.~(\ref{eq:zetadef})--(\ref{eq:phi1def}). 
Therefore, Eq.~(\ref{eq:vresult}) indicates that $v > 0$ for $\Delta\mu > 0$, as expected. 

The result for $D$ can be expressed as  
\begin{equation}
\label{eq:dresult}
	D = \lambda l + v\int_{-\infty}^\infty W(x)\,dx
\end{equation}
with the constant $\lambda$ and function $W(x)$ given below. 
The function $W(x)$ is defined by 
\begin{equation}
\label{eq:wdef}
	W(x) \equiv h(x) - \int_{-\infty}^xP(y)\,dy,  
\end{equation}
where the function $h(x)$ is defined by
\begin{equation}
\label{eq:hdef}
	h(x) \equiv \begin{cases}
		0 & (x < a_{-}), \\
		(x - a_{-})/l & (a_{-} \le x \le a_{+}), \\
		1 & (x > a_{+}),   
	\end{cases}
\end{equation}
and the rescaled steady-state distribution $P(x)$ is given by 
\begin{equation}
\label{eq:px}
	P(x) = g(x)\exp[-U(x)]
\end{equation}
with 
\begin{equation}
\label{eq:gx}
	g(x) = \begin{cases}
		C_{-} & (x < a_{-}), 
		\\
		\displaystyle 
		C_{-} - \frac{v}{D_0l}\int_{a_{-}}^x e^{U(y)}\,dy & (a_{-} \le x \le a_{+}), 
		\\
		C_{+} & (x > a_{+}). 
	\end{cases}
\end{equation}
Here, the constants $C_\pm$ are defined by
\begin{equation}
\label{eq:cpm}
	C_\pm \equiv \frac{(1 + \phi_0 u_\mp/D_0)v}{(u_{+} - u_{-})l}.
\end{equation}
The constant $\lambda$ in Eq.~(\ref{eq:dresult}) is given by 
\begin{equation}
\label{eq:lambda}
	\lambda \equiv - \frac{\zeta \psi_0u_{-} + \psi_1(u_{+} - u_{-})}
	{\zeta(D_0 + \phi_0u_{-}) + \phi_1(u_{+} - u_{-})}, 
\end{equation}
where $\psi_0$ and $\psi_1$ are integrals
\begin{align}
	\psi_0 &\equiv \int_{a_{-}}^{a_{+}} R(x)e^{U(x)}\,dx, 
	\label{eq:psi0def}
	\\
	\psi_1 &\equiv \int_{-\infty}^\infty e^{-U(x)}\,dx\int_x^{a_{+}} R(y)e^{U(y)}\,dy,
	\label{eq:psi1def}
\end{align}	
involving a new function $R(x)$ defined by
\begin{equation}
\label{eq:rdef}
	R(x) \equiv vW(x) - D_0P(x).   
\end{equation}

\section{Model with piecewise-linear potentials}
\label{sec:plinear}

As a specific example, we consider a model with piecewise linear potentials 
for which $V_0(x)$ is given by 
\begin{equation}
\label{eq:v0pl}
	V_0(x) = K\left|x\right|
\end{equation}
with a positive parameter $K$; see Fig.~\ref{fig:model}(b). 
The parameter for the locations of transitions is set as $a_{+} = 0$ 
(which implies $a_{-} = -l$). 
For this model the integrals needed to calculate the velocity and diffusion coefficient can be 
carried out analytically, as explained below.  

\subsection{Results}
\label{sec:results}

It is straightforward to obtain 
the constants involved in Eq.~(\ref{eq:vresult}) for $v$; 
the results are as follows: 
\begin{align}
\label{eq:zeta1}
	&\zeta = 2/\kappa, 
	\quad
	\phi_0 = \left(e^{\kappa l} - 1\right)/\kappa, 
	\quad
	\phi_1 = l/\kappa, 
	\\
\label{eq:upm1}
	&u_{+} = \omega_{+}, 
	\quad
	u_{-} = \omega_{-}e^{-\kappa l}, 
\end{align}
where 
\begin{equation}
\label{eq:kappa}
	\kappa \equiv K/k_\text{B}T. 
\end{equation}
Substituting expressions in Eqs.~(\ref{eq:zeta1}) and (\ref{eq:upm1}) into Eq.~(\ref{eq:vresult}), 
we obtain the average velocity $v$ of the motor for this model:
\begin{equation}
\label{eq:v1}
	v = \frac{\kappa D_0\left(\omega_{+} - \omega_{-}e^{-\kappa l}\right)}
	{\omega_{+} + (2/\kappa l)(\omega_{-} + \kappa D_0) 
	- (1 + 2/\kappa l)\omega_{-}e^{-\kappa l}}.
\end{equation}
 Note that $v$ monotonically increases with $\omega_{+}$, 
which is proportional to the forward transition rate $w_n^{+}(x)$, Eq.~(\ref{eq:wnpm}), 
and tends to the limiting value $\kappa D_0 = K/\gamma$. 
This is identical to the average velocity of a particle subject to a constant external force $K$. 
This is because, in the limit of large $\omega_{+}$, 
the particle stays only on the left-side slopes of potentials $V_n$ of Fig.~\ref{fig:model}(b), 
since the potential is switched from $V_n$ to $V_{n+1}$ 
right after the particle on the left-side slope of $V_n$ reaches 
the potential minimum at $x = nl$, 
and this switching brings the particle to the left-side slope of $V_{n+1}$, and so on. 

The expressions for the steady-state probability density $P(x)$ and 
the auxiliary function $W(x)$, both needed for calculating the diffusion coefficient,  
are presented in Appendix~\ref{sec:pxwx}. 
From the expressions for $W(x)$ given in Eqs.~(\ref{eq:w1}) and (\ref{eq:w2}), we have 
\begin{equation}
\label{eq:wint}
	\int_{-\infty}^\infty W(x)\,dx 
	= \frac{l}{2} - \frac{vl}{2\kappa D_0}\left(1 + \frac{2}{\kappa l}\right), 
\end{equation}
which contributes to the second term in Eq.~(\ref{eq:dresult}) for $D$. 
The integrals $\psi_0$ and $\psi_1$ defined by Eqs.~(\ref{eq:psi0def}) and (\ref{eq:psi1def}), 
respectively, can be carried out by substituting Eq.~(\ref{eq:rdef}) with $W(x)$ and $P(x)$ 
given in Eqs.~(\ref{eq:px1})--(\ref{eq:w2}) into these definitions to obtain 
\begin{align}
\label{eq:psi01}
	\psi_0 &= \frac{2v^2}{\kappa^2D_0}\left(1 + \frac{\kappa l}{4}  
	- \frac{e^{\kappa l} - 1}{\kappa l}\right) 
	- \frac{D_0\kappa l}{2},
	\\
\label{eq:psi11}
	\psi_1 &= \frac{v l}{2\kappa}\left(1 - \frac{4}{\kappa l}\right) - \frac{v^2l}{2\kappa^2D_0}. 
\end{align}
From these results and Eqs.~(\ref{eq:zeta1}) and (\ref{eq:upm1}), 
we find $\lambda$ defined by Eq.~(\ref{eq:lambda}), 
which contributes to the first term in Eq.~(\ref{eq:dresult}). 
Substituting this result for $\lambda$ together with Eq.~(\ref{eq:wint}) into Eq.~(\ref{eq:dresult}) 
provides us with the analytic expression for $D$. 

\begin{figure}[htb]
\includegraphics[scale = 0.9]{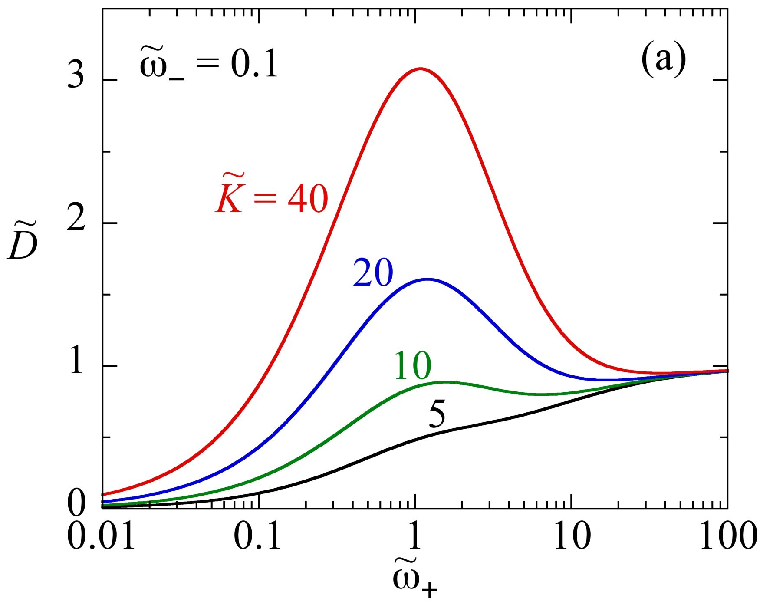}\\
\includegraphics[scale = 0.9]{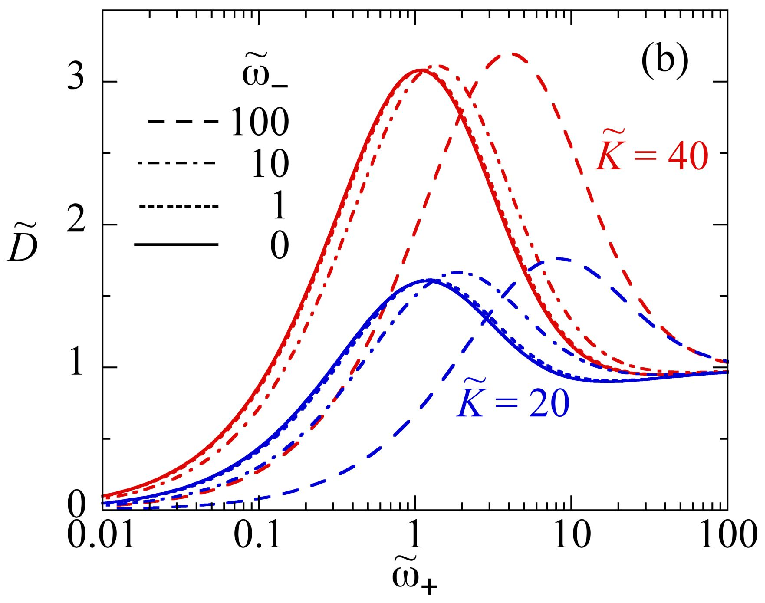}%
\caption{\label{fig:dw}%
	(Color online) 
	Dependence of the diffusion coefficient $\tilde D$ 
	on the forward transition rate $\tilde \omega_{+}$. 
	(a) The results for several choices of $\tilde K$ are shown in the case of $\tilde \omega_{-} = 0.1$. 
	(b) The results for several choices of the rate $\tilde \omega_{-}$ of backward transition, 
	in the cases of $\tilde K = 40$ and 20. 
}%
\end{figure}

The dependence of the diffusion coefficient 
on $\omega_{+}$ is shown in Fig.~\ref{fig:dw}(a)  
for several choices of $K$ and a fixed value of $\omega_{-}$. 
The result is represented in terms of dimensionless parameters defined by 
\begin{equation}
\label{eq:param}
	\tilde D \equiv D/D_0, 
	\quad
	\tilde \omega_\pm \equiv l\omega_\pm/D_0, 
	\quad
	\tilde K \equiv lK/k_\text{B}T.
\end{equation}
We observe that $\tilde D$ increases monotonically with $\tilde\omega_{+}$ for $\tilde K = 5$, 
whereas it has a peak around at $\tilde\omega_{+} \sim 1$ for large values of $\tilde K$. 
The peak height increases with $\tilde K$. 
In either case, $\tilde D$ tents to 1 ($D$ tends to $D_0$) from below as $\tilde\omega_{+} \to \infty$ 
(therefore, the curve $\tilde D(\tilde\omega_{+})$ exhibits a shallow dip when it has a peak). 
The reason why $D$ converges to $D_0$ is that, 
in this limiting case, the particle always experiences a constant external force 
as explained above, and hence its diffusion coefficient is the same as that of a free particle. 
The increase in the diffusion coefficient controlled by the transition rate shown in Fig.~\ref{fig:dw}(a) 
is the \textit{diffusion enhancement} in our model for molecular motors. 

Figure~\ref{fig:dw}(b) shows how the diffusion enhancement is affected by 
the rate parameter $\tilde\omega_{-}$ of the backward transition for the cases of 
$\tilde K = 20$ and 40. 
In the both cases, the peak height does not depend very much on $\tilde\omega_{-}$, 
while the peak position moves to the right (the direction of increasing $\tilde\omega_{+}$) 
as $\tilde\omega_{-}$ increases.

\subsection{Limit of small $\omega_{-}$}
\label{sec:limit}

We see from Fig.~\ref{fig:dw}(b) that, 
as $\tilde\omega_{-}$ decreases, $\tilde D$ as a function of $\tilde\omega_{+}$ for 
a given $\tilde K$ converges to a certain function [indicated by the solid line in Fig.~\ref{fig:dw}(b)]. 
This limiting function is obtained by setting $\omega_{-} = 0$ 
in the expression for $D$ obtained above; we have 
\begin{equation}
\label{eq:dlimit}
	D = \frac{v^3}{\kappa^3D_0^2}\left[1 + \frac{2D_0}{l\omega_{+}}
	+ 2\kappa l\left(\frac{D_0}{l\omega_{+}}\right)^2\right]
\end{equation}
with the limiting velocity
\begin{equation}
\label{eq:vlimit}
	v = \frac{\kappa D_0\omega_{+}}{\omega_{+} + 2D_0/l}, 
\end{equation}
which has the ``Michaelis--Menten type'' dependence~\cite{howard01} on $\omega_{+}$. 
We are interested in this limiting case, 
because the backward transition rate is negligibly small under the condition 
for the diffusion enhancement to be observed in the previous work~\cite{shinagawa16} 
for low ATP concentrations; 
the mechanism of the enhancement in this situation has not been clarified, 
as mentioned in the introduction. 

\begin{figure}[htb]
	\includegraphics[scale = 0.75]{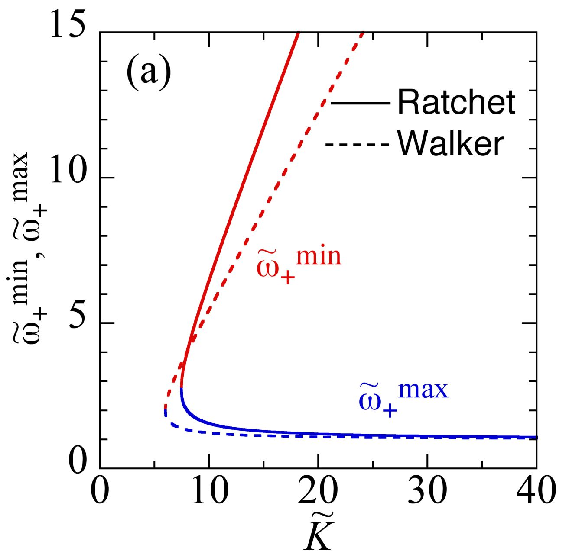}%
	\includegraphics[scale = 0.75]{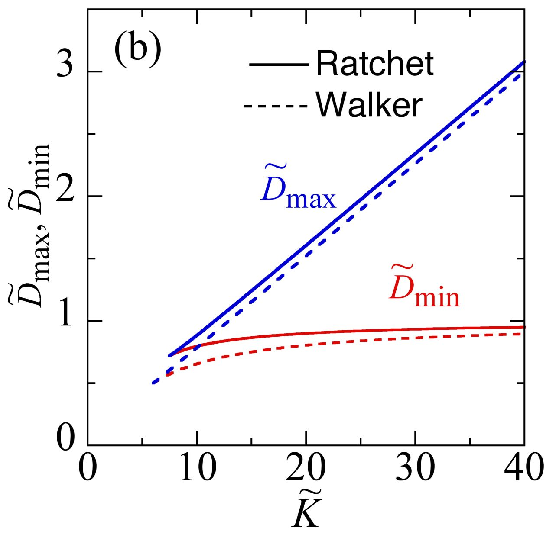}%
\caption{\label{fig:wdmm}%
	(Color online) 
	Dependence of (a) $\tilde\omega_{+}^\text{max}$ and $\tilde\omega_{+}^\text{min}$ 
	and (b) $\tilde D_\text{max}$ and $\tilde D_\text{min}$ on $\tilde K$.  
	Here, $\tilde\omega_{+}^\text{max}$ and $\tilde\omega_{+}^\text{min}$ are the positions of 
	the peak and the local minimum of function $\tilde D(\tilde\omega_{+})$, respectively, 
	and $\tilde D_\text{max}$ and $\tilde D_\text{min}$ are the values of $\tilde D$ at 
	these locations.  
	The solid lines are the results for the ratchet model given in Eq.~(\ref{eq:dlimit}) 
	in the limit of $\tilde\omega_{-} = 0$,  
	and the dashed lines are those of the extended Poisson walk, Eq.~(\ref{eq:dwalker}).
}%
\end{figure}

Equation~(\ref{eq:dlimit}) tells that if $\tilde K > 4 + 2\sqrt{3} \approx 7.46$ 
then function $\tilde D(\tilde\omega_{+})$ has a peak at 
\begin{equation}
\label{eq:wmax}
	\tilde\omega_{+}^\text{max} = \tilde K/2 - \sqrt{\tilde K^2/4 - 2\tilde K + 1}
\end{equation}
and a local minimum at $\tilde\omega_{+}^\text{min}$ give by Eq.~(\ref{eq:wmax}) 
with the sign of the last term being changed. 
The dependence of $\tilde\omega_{+}^\text{max}$ and  $\tilde\omega_{+}^\text{min}$ on $\tilde K$ 
are shown in Fig.~\ref{fig:wdmm}(a); 
and the peak height $\tilde D_\text{max}$ and 
the value of $\tilde D$ at the local minimum $\tilde D_\text{min}$ 
are plotted against $\tilde K$ in Fig.~\ref{fig:wdmm}(b). 
It is seen that 
the peak position $\tilde\omega_{+}^\text{max}$ tends to unity as
\begin{equation}
\label{eq:wmlim}
	\tilde\omega_{+}^\text{max} \approx 1 + 3/\tilde K
\end{equation}
in the large $\tilde K$ limit, 
whereas the peak height $\tilde D_\text{max}$ increases almost linearly in $\tilde K$. 
In fact, we have    
\begin{equation}
\label{eq:dmlim}
	\tilde D_\text{max} \approx 2\tilde K/27 + 1/9 
\end{equation}
for large $\tilde K$.  
Diffusion coefficient of more than twice that of free diffusion ($\tilde D > 2$) 
can be achieved for $\tilde K \gtrsim 25$. 
The mechanism of the diffusion enhancement in this limiting case is  
discussed in the next section.

\section{Extended Poisson walk}
\label{sec:pwalk}

\subsection{Diffusion of a random walker}

\begin{figure}[htb]
\includegraphics[scale = 0.6]{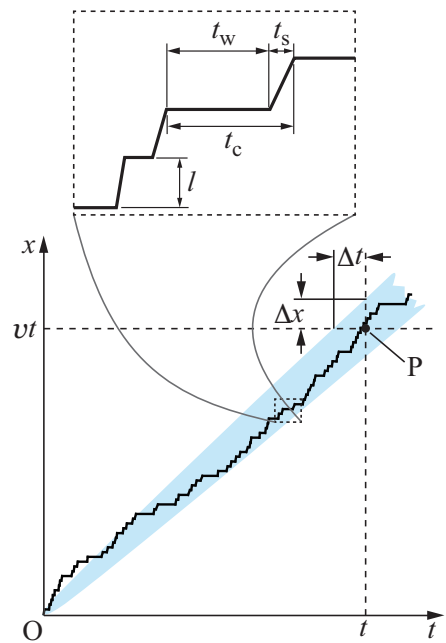}%
\caption{\label{fig:pwalk}%
	(Color online) 
	Trajectory of a random walker on a one-dimensional lattice of lattice constant $l$, 
	which mimics the trajectory of the particle in the ratchet model.  
	It stays on a lattice site waiting to step forward for a period of time $t_\text{w}$, 
	and it takes nonzero time $t_\text{s}$ to make a forward step. 
}%
\end{figure}

For qualitative understanding of the diffusion enhancement 
we have observed in the preceding section, 
let us take a look at the motion of the particle for the case 
in which the backward transition can be neglected (Section~IV B). 
The particle moves on the left-side slope of one of the V-shaped potentials 
shown in Fig.~\ref{fig:model} toward its bottom point right after a forward transition occurs.  
After reaching the bottom, 
it moves around the potential minimum until another forward transition occurs. 
Suppose that we plot the particle positions $x$ agains time $t$ 
at the occasions a forward transition occurs and 
the particle reaches the bottom of a potential for the first time. 
If we connect these points with line segments,  
we will have a trajectory like the one shown in Fig.~\ref{fig:pwalk}.

Such a trajectory may be viewed as a trajectory of a random walker 
on a one-dimensional lattice of lattice spacing $l$. 
The walker stays on a lattice site until it takes a forward step.  
The time $t_\text{w}$ for the walker to wait at the site (see the upper part of Fig.~\ref{fig:pwalk}) 
is a random variable. 
It also takes a nonzero time $t_\text{s}$ for the walker to move to the next lattice site; 
the stepping time $t_\text{s}$ is also a random variable. 
Let $t_\text{c}$ be the period of a cycle from the start of a waiting to 
the end of the stepping that follows it (Fig.~\ref{fig:pwalk}), i.e., 
$t_\text{c} = t_\text{w} + t_\text{s}$. 
The walker moves forward by distance $l$ every time it completes a cycle. 
Hence, the average velocity $v$ of the walker can be expressed as
\begin{equation}
\label{eq:vwalker}
	v = l/\tau_\text{c}  
\end{equation}
in terms of  the average $\tau_\text{c}$ of $t_\text{c}$. 

The diffusion coefficient of the walker may be obtained as follows.  
Consider a collection of trajectories of the walker starting from $x = 0$ at $t = 0$. 
The shaded area in Fig.~\ref{fig:pwalk} represents the region traversed by 
a large fraction of the trajectories. 
Let P the point on the line $x = vt$ that runs through the center of this region. 
Denoting the half-width of this region measured along the vertical line 
passing through this point by $\Delta x$ (see Fig.~\ref{fig:pwalk}), 
the diffusion coefficient can be estimated as 
\begin{equation}
\label{eq:dintuitive}
	D = \Delta x^2/2t = (v\Delta t)^2/2t,  
\end{equation}
where $t$ is the abscissa of point P and 
$\Delta t$ is the half-width of the shaded region in Fig.~\ref{fig:pwalk} 
measured along the horizontal line passing through point P. 
Now, $\Delta t$ can be calculated from the variance $\sigma_\text{c}^2$ of 
the cycle time $t_\text{c}$ of the walker as follows. 
The walker completes $N = vt/l$ cycles while it travels distance $vt$, 
which is the ordinate of point P. 
The variance of the time needed to complete $N$ cycles is $N\sigma_\text{c}^2$, 
and this variance is identified as $\Delta t^2$. 
Hence we have
\begin{equation}
\label{eq:dt2}
	\Delta t^2 = N\sigma_\text{c}^2 = vt\sigma_\text{c}^2/l.
\end{equation}
Substitution of this expression and Eq.~(\ref{eq:vwalker}) into Eq.~(\ref{eq:dintuitive}) results in  
\begin{equation}
\label{eq:dwalker}
	D = \sigma_\text{c}^2l^2/2\tau_\text{c}^3.
\end{equation} 
It should be noted that this expression for $D$ obtained by the qualitative arguments 
agrees exactly with the one derived by mathematically rigorous 
calculations~\cite{svoboda94, schnitzer95}; see also~\cite{reimann01, reimann02}.

Let us assume that the waiting of the walker is a Poisson process 
and hence the waiting time $t_\text{w}$  
is distributed exponentially as
\begin{equation}
\label{eq:wtime}
	f(t_\text{w}) = k\exp(-kt_\text{w}) 
\end{equation}
with a rate constant $k$, 
which is supposed to be related with the forward transition rate of our model of molecular motor.   
Then the average $\tau$ and the variance $\sigma^2$ of $t_\text{w}$ are given by 
$\tau = 1/k$ and $\sigma^2 = 1/k^2 = \tau^2$, respectively. 
If the stepping time $t_\text{s}$ is zero, the walker undergoes a Poisson random walk~\cite{vankampen07}. 
A walk with nonzero $t_\text{s}$ may be called an \textit{extended Poisson walk}. 
The average and variance of the stepping time $t_\text{s}$ will 
be denoted by $\tau_\text{s}$ and $\sigma_\text{s}^2$, respectively. 
Since the waiting and stepping are statistically independent, 
the average of $t_\text{c} = t_\text{w} + t_\text{s}$ are given as the sum of the 
averages of $t_\text{w}$ and $t_\text{s}$:  
$\tau_\text{c}= \tau + \tau_\text{s}$. 
Similarly, we have $\sigma_\text{c}^2 = \tau^2 + \sigma_\text{s}^2$. 
Hence, the expressions in Eqs.~(\ref{eq:vwalker})  and (\ref{eq:dwalker}) are rewritten as 
\begin{equation}
\label{eq:vdwalker}
	v = \frac{l}{\tau + \tau_\text{s}}, 
	\quad
	D = \frac{l^2(\tau^2 + \sigma_\text{s}^2)}{2(\tau + \tau_\text{s})^3}.
\end{equation}

Now, we examine the dependence of $D$ given in Eq.~(\ref{eq:vdwalker}) 
on the average waiting time $\tau$. 
If the waiting time is vanishingly small, the diffusion coefficient is determined by 
the stepping process, which yields
\begin{equation}
\label{eq:ds}
	D_\text{s} = \sigma_\text{s}^2l^2/2\tau_\text{s}^3. 
\end{equation}
As $\tau$ increases from zero, 
both the denominator and the numerator in the expression for $D$ 
in Eq.~(\ref{eq:vdwalker}) increase. 
It is easy to see that the increase in the denominator exceeds that in the numerator 
if $\tau$ is small enough or large enough,   
implying that $D$ decreases with increasing $\tau$ in these regions. 
On the other hand, if $\sigma_\text{s}$ is much smaller than $\tau_\text{s}$, 
then there is a region of $\tau$ where inequalities $\sigma_\text{s} \ll \tau \ll \tau_\text{s}$ hold. 
In this situation, we have significantly enhanced diffusion coefficient 
$D \sim D_\text{s}(\tau/\sigma_\text{s})^2$ compared with $D_\text{s}$. 

\begin{figure}[htb]
\includegraphics[scale = 0.9]{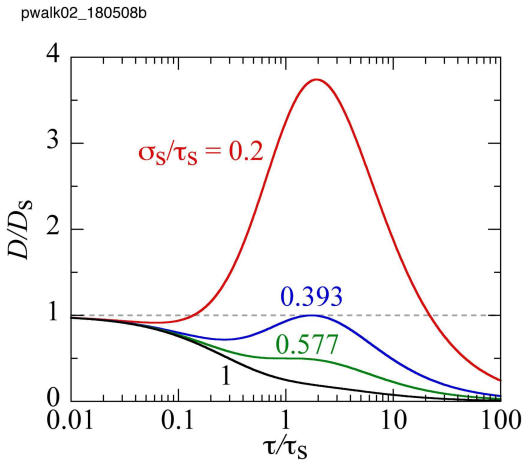}%
\caption{\label{fig:dpwalk}%
	(Color online) 
	Dependence of the diffusion coefficient $D$ given in Eq.~(\ref{eq:vdwalker}) 
	on the average waiting time $\tau$ for the extended Poisson walk. 
	The result is presented in dimensionless form for $\sigma_\text{s}/\tau_\text{s} = 1/5$, 
	$\big(2/\sqrt{3} - 1\big)^{1/2}$, $1/\sqrt{3}$, and 1.  
}%
\end{figure}

Precise calculations show that there is a region of $\tau$ 
where $D$ given in Eq.~(\ref{eq:vdwalker}) increases with $\tau$ 
if $\sigma_\text{s}/\tau_\text{s} < 1/\sqrt{3} \approx 0.577$, 
which indicates that function $D(\tau)$ has a peak, 
since $D$ decreases for large $\tau$ as explained above. 
The height of this peak exceeds $D_\text{s}$ (indicating diffusion enhancement) 
if $\sigma_\text{s}/\tau_\text{s} < \left(2/\sqrt{3} - 1\right)^{1/2} \approx 0.393$. 
These results are demonstrated in Fig.~\ref{fig:dpwalk}, 
where the dimensionless 
diffusion coefficient $D/D_\text{s}$ is plotted against the dimensionless waiting time $\tau/\tau_\text{s}$ 
for several choices of $\sigma_\text{s}/\tau_\text{s}$. 
As expected from the qualitative argument given in the preceding paragraph, 
we see that the diffusion coefficient is significantly enhanced 
for small enough $\sigma_\text{s}/\tau_\text{s}$ 
(see the graph of $\sigma_\text{s}/\tau_\text{s} = 0.2$). 
In the limit of small $\sigma_\text{s}/\tau_\text{s}$, 
the location $\tau_\text{max}$ and the height $D_\text{max}$ 
of the peak in $D(\tau)$ due to the diffusion enhancement 
can be estimated to be 
\begin{equation}
\label{eq:tmax}
	\tau_\text{max} \approx 2\tau_\text{s}, 
	\quad
	D_\text{max} \approx 2l^2/27\tau_\text{s} = (2\tau_\text{s}/3\sigma_\text{s})^2D_\text{s}/3,
\end{equation}
respectively, from Eq.~(\ref{eq:vdwalker}) by setting $\sigma_\text{s} \approx 0$. 
To summarize, 
\textit{%
the diffusion is enhanced when the waiting time is comparable to the 
stepping time for the extended Poisson walk if the fluctuation of the stepping time is small enough. 
}%

\subsection{Ratchet as a random walker}

Now, we discuss the correspondence between the extended Poisson walker 
and the ratchet model studied in the preceding section. 
Let $t$ be the time needed for a particle moving 
in potential $V_0(x)$ given by Eq.~(\ref{eq:v0pl}) to arrive at $x = 0$ for the first time 
provided that it has started at $x = -l$. 
The probability density function $p(t)$ of $t$ (the first-passage time) is given~\cite{hu10} by 
\begin{equation}
\label{eq:pt}
	p(t) = \frac{l}{\sqrt{4\pi D_0 t^3}}\exp\left[-\frac{(l - \kappa D_0t)^2}{4D_0t}\right].
\end{equation}
The average $\tau_\text{r}$ and the variance $\sigma^2_\text{r}$ of $t$ are calculated from 
this distribution as
\begin{equation}
\label{eq:taua}
	\tau_\text{r} = l/\kappa D_0, 
	\quad
	\sigma^2_\text{r} = 2l/\kappa^3D_0^2. 
\end{equation}
Note that the same results can be obtained from the closed-form formulas 
for the moments of the first-passage time; 
see Ref.~\cite{reimann02} and Sec.~7 in Ref.~\cite{hanggi90}.  
It seems reasonable to identify the first-passage time 
with the stepping time $t_\text{s}$ of the walker. 
Therefore, $\tau_\text{s}$ and $\sigma_\text{s}$ of the walker should 
correspond to $\tau_\text{r}$ and $\sigma_\text{r}$ of the ratchet. 
The rate $k$ associated with the waiting time of the walker should correspond to 
the rate of the transition in the ratchet. 
If the thermal equilibrium of $x$ is achieved before the transition, 
then this rate is estimated as 
\begin{equation}
\label{eq:mrate}
	w = \int_{-\infty}^\infty w_0^{+}(x) P_\text{eq}(x)\,dx 
	= \kappa\omega_{+}/2, 
\end{equation}
where $P_\text{eq}(x) \equiv (\kappa/2)\exp(-\kappa|x|)$ is the equilibrium
distribution for the position of a particle in potential $V_0(x)$. 
Identifying $\tau_\text{r}$, $\sigma_\text{r}^2$, and $w$ of the ratchet 
with $\tau_\text{s}$, $\sigma_\text{s}$, and $k = 1/\tau$ of the walker, respectively, 
we obtain the same expression for $v$ as Eq.~(\ref{eq:vlimit}) and
\begin{equation}
\label{eq:dwalk}
	D = \frac{v^3}{\kappa^3D_0^2}\left[1 + 2\kappa l\left(\frac{D_0}{l\omega_{+}}\right)^2\right]
\end{equation}
from the results~(\ref{eq:vdwalker}) for the walker. 
Equation~(\ref{eq:dwalk}) agrees with Eq.~(\ref{eq:dlimit}) except the second term 
in the brackets in the latter, which is absent in the former. 

\begin{figure}[htb]
\includegraphics[scale = 0.9]{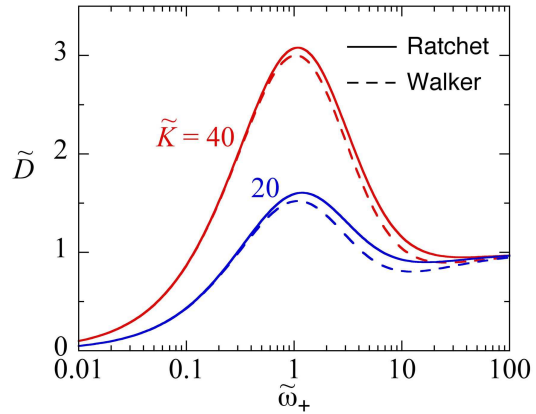}%
\caption{\label{fig:dww}%
	(Color online) 
	Dependence of the diffusion coefficient $\tilde D$ on the rate parameter $\tilde\omega_{+}$. 
	The result for the ratchet model in the absence of the backward transition, Eq.~(\ref{eq:dlimit}), 
	is shown by the solid line, and that for for the extended Poisson walk, Eq.~(\ref{eq:dwalk}),  
	by the dashed line. 
	The cases of $\tilde K = 20$ and 40 are presented.  
	The solid lines here are identical to those in Fig.~\ref{fig:dw}(b). 
}%
\end{figure}

The diffusion coefficient~(\ref{eq:dwalk}) obtained for the extended Poisson walk 
is plotted, in the dimensionless form,  against $\tilde\omega_{+} = l\omega_{+}/D_0$ in Fig.~\ref{fig:dww}, 
together with the result of Eq.~(\ref{eq:dlimit}) for the ratchet model. 
We see that the position and the height of the peak of function $\tilde D(\tilde\omega_{+})$ 
for the ratchet model agree reasonably well with those of the extended Poisson walk. 
The peak for the latter model is located at $\tilde\omega_{+} = \tilde\omega_{+}^\text{max} 
\equiv [\tilde K - (\tilde K^2 - 6\tilde K)^{1/2}]/3$ for $\tilde K > 6$; 
the peak position approaches unity as $\tilde\omega_{+}^\text{max} \approx 1 + 3/2\tilde K$ 
in the limit of large $\tilde K$, whereas the peak height increases linearly in $\tilde K$ 
as $\tilde D_\text{max} \approx 2\tilde K/27 + 1/27$ in the same limit. 
These expressions are to be compared with Eqs.~(\ref{eq:wmlim}) and (\ref{eq:dmlim}), respectively. 
The dependences of $\tilde\omega_{+}^\text{max}$ and $\tilde D_\text{max}$, 
as well as $\tilde\omega_{+}^\text{min}$ and $\tilde D_\text{min}$ 
[which are the values of $\tilde\omega_{+}$ and $\tilde D$ 
at the local minimum of $\tilde D(\tilde\omega_{+})$], 
on $\tilde K$ are shown in Fig.~\ref{fig:wdmm}(a) and (b), respectively.

These results demonstrate that the pronounced enhancement of diffusion 
observed in the ratchet model for large $\tilde K$ is well described by the extended Poisson walk. 
Hence, we suggest that the mechanism for  
the diffusion enhancement observed in our model of molecular motor 
is essentially the same as that for the extended Poisson walk; 
\textit{%
the diffusion is enhanced when the waiting time for the transition 
is comparable to the time for the particle to slide down the potential slope. 
}%
Remember that the condition for the diffusion enhancement to occur 
in the extended Poisson walk is that the fluctuation (the standard deviation) of 
the stepping time $t_\text{s}$ should be somewhat smaller than its average. 
This condition is satisfied for large $\tilde K$ in the ratchet model, 
since we have $(\sigma_\text{r}/\tau_\text{r})^2 = 2/\kappa l = 2/\tilde K$, 
as can be seen from Eq.~(\ref{eq:taua}). 
This explains why the larger $\tilde K$ is, the more salient the diffusion enhancement is. 

\subsection{A previous model of molecular motors}

Let us see if the analogy between a ratchet model and the extended Poisson walk will work for 
the diffusion enhancement in the previous model of molecular motors~\cite{shinagawa16}, 
which has some relevance to the F$_1$-ATPase~\cite{kawaguchi14}. 
This model is also a potential-switching ratchet described by the Fokker-Planck equation~(\ref{eq:fpeq}), 
but with different functions for the potential and the transition rates:  
the potential for and the forward transition rate from ``state~0'' are given by $V_0(x) = Kx^2/2$ and 
$w_0^{+}(x) = k\exp(ax)$, respectively, 
and a constant external force $F$ is applied, 
where $K$, $k$, and $a$ are positive constants  
($k$ is proportional to the ATP concentration). 
The particle subject to the potential $V_0$ and the external force 
is in mechanical equilibrium at $x = F/K$. 
Therefore, the average waiting time for the transition from state~0 to state~1 can 
roughly be estimated as $\tau \sim 1/w_0^{+}(F/K) = (1/k)\exp(-aF/K)$. 
Now, it will take about $\tau_\text{s} \sim \gamma/K = k_\text{B}T/D_0K$ 
for the particle to slide down on the potential $V_1(x)$ from $x = F/K$ to 
the vicinity of $x = F/K + l$, the mechanical-equilibrium position in state~1,  
after the transition. 
Then, Eq.~(\ref{eq:tmax}) predicts that the diffusion coefficient 
as a function of $F$ for the motor will be maximized at 
$F = F_\text{max} \sim (K/a)\ln(k_\text{B}T/2D_0kK)$ 
with maximum value $D_\text{max} \sim 2l^2D_0K/27k_\text{B}T$. 
These dependences of $F_\text{max}$ and $D_\text{max}$ on $K$ and $k$ agree within numerical factors 
with those reported in Ref.~\cite{shinagawa16} for large $K$ and small $k$ 
(low ATP concentration). 
Hence, we think that the mechanism of the diffusion enhancement in molecular motors 
under low ATP concentrations reported in our previous work~\cite{shinagawa16} is 
essentially the same as that in the extended Poisson walk.

\section{Concluding Remarks}
\label{sec:conclusion}

We have proposed a solvable model of ratchet type for the Brownian motor 
to elucidate the mechanism underlying the diffusion enhancement reported 
for a model of molecular motors in our previous work~\cite{shinagawa16}. 
We have suggested that the diffusion enhancement of the present model 
observed in a certain range of the transition rate (Sec.~\ref{sec:plinear}) 
and of the previous model~\cite{shinagawa16} for certain range of external force 
at low ATP concentrations 
is essentially the same as the one for a simpler system 
which we call the extended Poisson walk (Sec.~\ref{sec:pwalk}). 
In this random walk on a one-dimensional lattice, 
each step of the walk consists of two processes. 
One is the Poisson process for the walker to wait on a lattice site, 
and the other is the stepping to the next site that takes a nonzero time. 
The enhancement of diffusion occurs when the average waiting time is 
comparable to the stepping time. 
In the ratchet model, the waiting time corresponds to 
the time for the Brownian particle to stay in one potential, 
and the stepping corresponds to the sliding of the particle on the potential slope 
after a chemical transition is made. 

The analogy between the extended Poisson walk and the ratchet model studied in Sec.~\ref{sec:plinear} 
works only in the limit of small $\omega_{-}$ (the rate of backward transition). 
Therefore this analogy cannot explain the result presented in Fig.~\ref{fig:dw}(b) 
that the peak position of 
the diffusion coefficient as a function of the forward transition rate moves 
to the right as $\omega_{-}$ increases. 
Whether this behavior can be understood on the basis of a simple physical picture 
will be investigated in a future work.

\appendix

\section{Derivation of Eqs.~(\ref{eq:vresult}) and (\ref{eq:dresult})}
\label{sec:derivation}

To calculate the velocity from the formula~(\ref{eq:vformula}), 
we need to obtain the steady-state solutions $P_n(x)$ 
to the Fokker-Planck equations~(\ref{eq:fpeq}).   
Let $J(x)$ be the function defined by 
\begin{equation}
\label{eq:jdef}
	J(x) \equiv \left[f(x) - D_0\frac{d}{dx}\right]P(x),
\end{equation}
where $P(x)$ is the rescaled $P_0(x)$ introduced in Sec.~\ref{sec:retchet}. 
Making use of the relation $P_n(x) = P_0(x - nl)$ and Eq.~(\ref{eq:wnpm}), 
we obtain 
\begin{equation}
\label{eq:jeq}
	\frac{dJ}{dx} = \mathcal{J}\left[\delta(x - a_{-}) - \delta(x - a_{+})\right] 
\end{equation} 
with
\begin{equation}
\label{eq:jc}
	\mathcal{J} \equiv \omega_{+}P(a_{+}) - \omega_{-}P(a_{-}) 
\end{equation}
from Eq.~(\ref{eq:fpeq}). 
Equation~(\ref{eq:jeq}) implies that $J(x)$ is piecewise constant, 
and the boundary condition $P(x) \to 0$ as $x \to \pm\infty$ and Eq.~(\ref{eq:jdef}) 
suggest that $J(x) \to 0$ as $x \to \pm\infty$. 
Therefore, Eq.~(\ref{eq:jeq}) is integrated to yield 
\begin{equation}
\label{eq:jx}
	J(x) = \mathcal{J}\theta(x - a_{-})\theta(a_{+} - x),
\end{equation}
where $\theta(x)$ is the step function: 
$\theta(x) = 0$ for $x < 0$ and $\theta(x) = 1$ for $x \ge 0$. 
By making use of Eqs.~(\ref{eq:jdef}) and (\ref{eq:jx}), 
we can rewrite Eq.~(\ref{eq:vformula}) as 
\begin{equation}
\label{eq:vj}
	v = \int_{-\infty}^\infty J(x)\,dx = \mathcal{J}l.
\end{equation}

Substituting Eq.~(\ref{eq:jx}) into Eq.~(\ref{eq:jdef}) 
and integrating the resulting equation with the boundary condition $P \to 0$ as $|x| \to \infty$, 
we obtain 
\begin{equation}
\label{eq:pxa}
	P(x) = \begin{cases}
		C_{-}\exp[-U(x)] & (x < a_{-}),
		\\
		\varphi(x)\exp[-U(x)] & (a_{-} \le x \le a_{+}), 
		\\
		C_{+}\exp[-U(x)] & (x > a_{+}),
	\end{cases}
\end{equation}
where $U(x)$ is defined in Eq.~(\ref{eq:uxdef}), 
function $\varphi(x)$ is defined by 
\begin{equation}
\label{eq:phi}
	\varphi(x) = C_{-} - \frac{\mathcal{J}}{D_0}\int_{a_{-}}^x e^{U(y)}\,dy, 
\end{equation}
and constants $C_\mp$ by 
\begin{equation}
\label{eq:cmp}
	C_\mp = \frac{1 + \phi_0 u_\pm/D_0}{u_{+} - u_{-}}\mathcal{J} 
\end{equation}
with $\phi_0$ and $u_\pm$ defined in Eqs.~(\ref{eq:phi0def}) and (\ref{eq:upmdef}). 
We have also used Eq.~(\ref{eq:jc}) to get Eq.~(\ref{eq:cmp}). 
From Eqs.~(\ref{eq:vj})--(\ref{eq:cmp}) we obtain Eqs.~(\ref{eq:px})--(\ref{eq:cpm}). 

The expression~(\ref{eq:pxa}) for $P(x)$ together with Eqs.~(\ref{eq:phi}) and (\ref{eq:cmp}) 
contain the unknown constant $\mathcal{J}$. 
This constant can be determined from the normalization condition~(\ref{eq:pnorm}), 
which yields
\begin{equation}
\label{eq:jresult}
	\mathcal{J} = \frac{(u_{+} - u_{-})D_0}{\zeta(D_0 + \phi_0u_{-}) + \phi_1(u_{+} - u_{-})}, 
\end{equation}
where $\phi_1$ is defined in Eq.~(\ref{eq:phi1def}). 
From Eqs.~(\ref{eq:jresult}) and (\ref{eq:vj}) we obtain Eq.~(\ref{eq:vresult}). 
 
To calculate the diffusion coefficient from the formula~(\ref{eq:dformula}), 
we need to solve Eq.~(\ref{eq:qeq}) to obtain $Q(x)$. 
Let $L(x)$ be the function defined by 
\begin{equation}
\label{eq:ldef}
	L(x) \equiv \left[f(x) - D_0\frac{d}{dx}\right]Q(x). 
\end{equation}
Then, Eq.~(\ref{eq:qeq}) is rewritten as 
\begin{align}
	\frac{dL}{dx} &= \lambda\left[\delta(x - a_{-}) - \delta(x - a_{+})\right]
	\nonumber\\
	&\quad{}+ \left[f(x) - v - 2D_0\frac{d}{dx}\right]P(x)
\label{eq:leq}
\end{align}
with 
\begin{equation}
\label{eq:lambdadef}
	\lambda \equiv \omega_{+}Q(a_{+}) - \omega_{-}Q(a_{-}).  
\end{equation}
Taking account of the boundary condition $L(x) \to 0$ as $|x| \to \infty$, 
which comes from the similar condition for $Q(x)$ and Eq.~(\ref{eq:ldef}), 
we integrate Eq.~(\ref{eq:leq}) to get 
\begin{equation}
\label{eq:lresult}
	L(x) = R(x) + \lambda\theta(x - a_{-})\theta(a_{+} - x),
\end{equation}
where $R(x)$ is defined by
\begin{equation}
\label{eq:rdefa}
	R(x) \equiv \int_{-\infty}^x \left[f(y) - v\right]P(y)\,dy - 2D_0P(x).  
\end{equation}
It is convenient to rewrite this expression as follows. 
From the definition~(\ref{eq:vformula}) of $f(x)$ and the expression~(\ref{eq:px}) for $P(x)$, 
we have 
\begin{equation}
\label{eq:d3}
	\int_{-\infty}^x f(y)P(y)\,dy = D_0P(x) - D_0\int_{-\infty}^x \frac{dg(y)}{dy}e^{-U(y)}\,dy
\end{equation}
by making use of  integration by parts. 
Now, the expression~(\ref{eq:gx}) for $g(x)$ is used 
to rewrite the last term in Eq.~(\ref{eq:d3}) as 
\begin{equation}
\label{eq:d4}
	D_0\int_{-\infty}^x \frac{dg(y)}{dy}e^{-U(y)}\,dy = -vh(x), 
\end{equation}
where $h(x)$ is defined in Eq.~(\ref{eq:hdef}). 
Substitution of Eq.~(\ref{eq:d3}) with Eq.~(\ref{eq:d4}) into Eq.~(\ref{eq:rdefa}) leads to  
Eq.(\ref{eq:rdef}), i.e., 
\begin{equation}
\label{eq:rx}
	R(x) = vW(x) - D_0P(x)
\end{equation}
with $W(x)$ defined in Eq.~(\ref{eq:wdef}). 

Having the function $L(x)$ thus obtained, 
we substitute Eq.~(\ref{eq:lresult}) with Eq.~(\ref{eq:rx}) into Eq.~(\ref{eq:ldef}) 
and integrate the resulting equation to get  
\begin{equation}
\label{eq:qx}
	Q(x) = e^{-U(x)}\left[C - \frac{1}{D_0}\int_{a_{-}}^x L(y)e^{U(y)}\,dy\right], 
\end{equation}
where constant $C$ is given by
\begin{equation}
\label{eq:cdef}
	C = \frac{(D_0 + \phi_0 u_{+})\lambda + \psi_0 u_{+}}{(u_{+} - u_{-})D_0}
\end{equation}
with $\phi_0$, $u_\pm$, and $\psi_0$ defined by Eqs.~(\ref{eq:phi0def}), (\ref{eq:upmdef}), 
and (\ref{eq:psi0def}), respectively. 
We have also used Eq.~(\ref{eq:lambdadef}) to get Eq.~(\ref{eq:cdef}). 

The expression~(\ref{eq:qx}) contains the unknown constant $\lambda$ through $C$ 
given in Eq.~(\ref{eq:cdef}).  
This constant cannot be determined uniquely, 
because $Q(x)$ defined as the solution of Eq.~(\ref{eq:qeq}) has ambiguity: 
if $Q(x)$ is a solution of this equation, 
then $Q(x) + cP(x)$ with $c$ an arbitrary constant is also a solution.  
However, this ambiguity does not affect the right-hand side of formula~(\ref{eq:dformula}), 
since we have 
\[
	\int_{-\infty}^\infty \left[f(x) - v\right]P(x)\,dx = 0
\] 
because of the first equation in Eq.~(\ref{eq:vformula}). 
Therefore, we can assign any value to $\lambda$. 
We find it convenient to determine $\lambda$ from the condition 
\begin{equation}
\label{eq:qcond}
	\int_{-\infty}^\infty Q(x)\,dx = 0, 
\end{equation}
from which we obtain Eq.~(\ref{eq:lambda}). 

Now, we have everything we need to calculate the diffusion coefficient 
by using the formula~(\ref{eq:dformula}), 
which reads 
\begin{equation}
\label{eq:d1}
	D = D_0 + \int_{-\infty}^\infty f(x)Q(x)\,dx 
\end{equation}
due to Eq.~(\ref{eq:qcond}).  
We will rewrite the second term of Eq.~(\ref{eq:d1}), 
because the expression for $Q(x)$ given in Eq.~(\ref{eq:qx}) is quite complicated 
and hence Eq.~(\ref{eq:d1}) is not convenient for practical use. 
First, we use Eqs.~(\ref{eq:ldef}) and (\ref{eq:lresult}) to proceed as  
\begin{equation*}
	\int_{-\infty}^\infty f(x)Q(x)\,dx = \int_{-\infty}^\infty L(x)\,dx
	= \lambda l + \int_{-\infty}^\infty R(x)\,dx.
\end{equation*}
Next, we use Eq.~(\ref{eq:rx}) to obtain Eq.~(\ref{eq:dresult}).

\section{Expressions for $P(x)$ and $W(x)$}
\label{sec:pxwx}

Here, we present the explicit expressions for $P(x)$ and $W(x)$ 
obtained for the model considered in Sec.~\ref{sec:plinear}.  
The steady-state probability density $P(x)$ in state 0 is found to be 
\begin{equation}
\label{eq:px1}
	P(x) = g(x) e^{-\kappa |x|}, 
\end{equation}
where $g(x) = C_{-}$ for $x < -l$, 
\begin{equation}
\label{eq:gx1}
	g(x) = C_{-} - \frac{v}{\kappa lD_0}\left(e^{\kappa l} - e^{-\kappa x}\right)
\end{equation}
for $-l \le x \le 0$, and $g(x) = C_{+}$ for $x > 0$ with 
$C_\pm$ obtained by substituting Eqs.~(\ref{eq:zeta1}) and (\ref{eq:upm1}) into Eq.~(\ref{eq:cpm}). 

The function $W(x)$ defined by Eq.~(\ref{eq:wdef}) is given by 
\begin{equation}
\label{eq:w1}
	W(x) = \begin{cases}
		-(C_{-}/\kappa)e^{\kappa x} & (x < -l) 
		\\(C_{+}/\kappa)e^{-\kappa x} & (x > 0)
	\end{cases}
\end{equation}
and 
\begin{equation}
\label{eq:w2}
	W(x) = C_{+}\zeta\left(1 + \frac{x}{l}\right)
	 - \frac{C_{-}}{\kappa}e^{\kappa x} 
	 + \frac{v}{\kappa^2lD_0}\left(e^{\kappa(x + l)} - 1\right)
\end{equation}
for $-l \le x \le 0$.

\begin{acknowledgments}
This work was supported in part by JSPS KAKENHI Grant Number JP17K05562 (KS) 
and by the Research Complex Promotion Program (RK).
\end{acknowledgments}

\bibliography{ExactDE}

\begin{thebibliography}{20}%
\makeatletter
\providecommand \@ifxundefined [1]{%
 \@ifx{#1\undefined}
}%
\providecommand \@ifnum [1]{%
 \ifnum #1\expandafter \@firstoftwo
 \else \expandafter \@secondoftwo
 \fi
}%
\providecommand \@ifx [1]{%
 \ifx #1\expandafter \@firstoftwo
 \else \expandafter \@secondoftwo
 \fi
}%
\providecommand \natexlab [1]{#1}%
\providecommand \enquote  [1]{``#1''}%
\providecommand \bibnamefont  [1]{#1}%
\providecommand \bibfnamefont [1]{#1}%
\providecommand \citenamefont [1]{#1}%
\providecommand \href@noop [0]{\@secondoftwo}%
\providecommand \href [0]{\begingroup \@sanitize@url \@href}%
\providecommand \@href[1]{\@@startlink{#1}\@@href}%
\providecommand \@@href[1]{\endgroup#1\@@endlink}%
\providecommand \@sanitize@url [0]{\catcode `\\12\catcode `\$12\catcode
  `\&12\catcode `\#12\catcode `\^12\catcode `\_12\catcode `\%12\relax}%
\providecommand \@@startlink[1]{}%
\providecommand \@@endlink[0]{}%
\providecommand \url  [0]{\begingroup\@sanitize@url \@url }%
\providecommand \@url [1]{\endgroup\@href {#1}{\urlprefix }}%
\providecommand \urlprefix  [0]{URL }%
\providecommand \Eprint [0]{\href }%
\providecommand \doibase [0]{http://dx.doi.org/}%
\providecommand \selectlanguage [0]{\@gobble}%
\providecommand \bibinfo  [0]{\@secondoftwo}%
\providecommand \bibfield  [0]{\@secondoftwo}%
\providecommand \translation [1]{[#1]}%
\providecommand \BibitemOpen [0]{}%
\providecommand \bibitemStop [0]{}%
\providecommand \bibitemNoStop [0]{.\EOS\space}%
\providecommand \EOS [0]{\spacefactor3000\relax}%
\providecommand \BibitemShut  [1]{\csname bibitem#1\endcsname}%
\let\auto@bib@innerbib\@empty
\bibitem [{\citenamefont {Berg}(1993)}]{berg93}%
  \BibitemOpen
  \bibfield  {author} {\bibinfo {author} {\bibfnamefont {H.~C.}\ \bibnamefont
  {Berg}},\ }\href@noop {} {\emph {\bibinfo {title} {Random Walks in
  Biology}}}\ (\bibinfo  {publisher} {Princeton University Press},\ \bibinfo
  {year} {1993})\ Chap.~\bibinfo {chapter} {6}\BibitemShut {NoStop}%
\bibitem [{\citenamefont {Costantini}\ and\ \citenamefont
  {Marchesoni}(1999)}]{costantini99}%
  \BibitemOpen
  \bibfield  {author} {\bibinfo {author} {\bibfnamefont {G.}~\bibnamefont
  {Costantini}}\ and\ \bibinfo {author} {\bibfnamefont {F.}~\bibnamefont
  {Marchesoni}},\ }\href@noop {} {\bibfield  {journal} {\bibinfo  {journal}
  {Europhys.\ Lett.}\ }\textbf {\bibinfo {volume} {48}},\ \bibinfo {pages}
  {491} (\bibinfo {year} {1999})}\BibitemShut {NoStop}%
\bibitem [{\citenamefont {Reimann}\ \emph {et~al.}(2001)\citenamefont
  {Reimann}, \citenamefont {Van~den Broeck}, \citenamefont {Linke},
  \citenamefont {H\"{a}nggi}, \citenamefont {Rubi},\ and\ \citenamefont
  {P\'{e}rez-Madrid}}]{reimann01}%
  \BibitemOpen
  \bibfield  {author} {\bibinfo {author} {\bibfnamefont {P.}~\bibnamefont
  {Reimann}}, \bibinfo {author} {\bibfnamefont {C.}~\bibnamefont {Van~den
  Broeck}}, \bibinfo {author} {\bibfnamefont {H.}~\bibnamefont {Linke}},
  \bibinfo {author} {\bibfnamefont {P.}~\bibnamefont {H\"{a}nggi}}, \bibinfo
  {author} {\bibfnamefont {J.~M.}\ \bibnamefont {Rubi}}, \ and\ \bibinfo
  {author} {\bibfnamefont {A.}~\bibnamefont {P\'{e}rez-Madrid}},\ }\href@noop
  {} {\bibfield  {journal} {\bibinfo  {journal} {Phys.\ Rev.\ Lett.}\ }\textbf
  {\bibinfo {volume} {87}},\ \bibinfo {pages} {010602} (\bibinfo {year}
  {2001})}\BibitemShut {NoStop}%
\bibitem [{\citenamefont {Reimann}\ \emph {et~al.}(2002)\citenamefont
  {Reimann}, \citenamefont {Van~den Broeck}, \citenamefont {Linke},
  \citenamefont {H\"{a}nggi}, \citenamefont {Rubi},\ and\ \citenamefont
  {P\'{e}rez-Madrid}}]{reimann02}%
  \BibitemOpen
  \bibfield  {author} {\bibinfo {author} {\bibfnamefont {P.}~\bibnamefont
  {Reimann}}, \bibinfo {author} {\bibfnamefont {C.}~\bibnamefont {Van~den
  Broeck}}, \bibinfo {author} {\bibfnamefont {H.}~\bibnamefont {Linke}},
  \bibinfo {author} {\bibfnamefont {P.}~\bibnamefont {H\"{a}nggi}}, \bibinfo
  {author} {\bibfnamefont {J.~M.}\ \bibnamefont {Rubi}}, \ and\ \bibinfo
  {author} {\bibfnamefont {A.}~\bibnamefont {P\'{e}rez-Madrid}},\ }\href@noop
  {} {\bibfield  {journal} {\bibinfo  {journal} {Phys.\ Rev.\ E}\ }\textbf
  {\bibinfo {volume} {65}},\ \bibinfo {pages} {031104} (\bibinfo {year}
  {2002})}\BibitemShut {NoStop}%
\bibitem [{\citenamefont {Speer}\ \emph {et~al.}(2012)\citenamefont {Speer},
  \citenamefont {Eichhorn},\ and\ \citenamefont {Reimann}}]{speer12}%
  \BibitemOpen
  \bibfield  {author} {\bibinfo {author} {\bibfnamefont {D.}~\bibnamefont
  {Speer}}, \bibinfo {author} {\bibfnamefont {R.}~\bibnamefont {Eichhorn}}, \
  and\ \bibinfo {author} {\bibfnamefont {P.}~\bibnamefont {Reimann}},\
  }\href@noop {} {\bibfield  {journal} {\bibinfo  {journal} {EPL}\ }\textbf
  {\bibinfo {volume} {85}},\ \bibinfo {pages} {60004} (\bibinfo {year}
  {2012})}\BibitemShut {NoStop}%
\bibitem [{\citenamefont {Hayashi}\ \emph {et~al.}(2015)\citenamefont
  {Hayashi}, \citenamefont {Sasaki}, \citenamefont {Nakamura}, \citenamefont
  {Kudo}, \citenamefont {Inoue}, \citenamefont {Noji},\ and\ \citenamefont
  {Hayashi}}]{hayashi15}%
  \BibitemOpen
  \bibfield  {author} {\bibinfo {author} {\bibfnamefont {R.}~\bibnamefont
  {Hayashi}}, \bibinfo {author} {\bibfnamefont {K.}~\bibnamefont {Sasaki}},
  \bibinfo {author} {\bibfnamefont {S.}~\bibnamefont {Nakamura}}, \bibinfo
  {author} {\bibfnamefont {S.}~\bibnamefont {Kudo}}, \bibinfo {author}
  {\bibfnamefont {Y.}~\bibnamefont {Inoue}}, \bibinfo {author} {\bibfnamefont
  {H.}~\bibnamefont {Noji}}, \ and\ \bibinfo {author} {\bibfnamefont
  {K.}~\bibnamefont {Hayashi}},\ }\href@noop {} {\bibfield  {journal} {\bibinfo
   {journal} {Phys.\ Rev.\ Lett}\ }\textbf {\bibinfo {volume} {114}},\ \bibinfo
  {pages} {248101} (\bibinfo {year} {2015})}\BibitemShut {NoStop}%
\bibitem [{\citenamefont {Kim}\ \emph {et~al.}(2017)\citenamefont {Kim},
  \citenamefont {Bowman}, \citenamefont {Del Bonis-O'Donnel}, \citenamefont
  {Matzavions},\ and\ \citenamefont {Stein}}]{kim17}%
  \BibitemOpen
  \bibfield  {author} {\bibinfo {author} {\bibfnamefont {D.}~\bibnamefont
  {Kim}}, \bibinfo {author} {\bibfnamefont {C.}~\bibnamefont {Bowman}},
  \bibinfo {author} {\bibfnamefont {J.~T.}\ \bibnamefont {Del Bonis-O'Donnel}},
  \bibinfo {author} {\bibfnamefont {A.}~\bibnamefont {Matzavions}}, \ and\
  \bibinfo {author} {\bibfnamefont {D.}~\bibnamefont {Stein}},\ }\href@noop {}
  {\bibfield  {journal} {\bibinfo  {journal} {Phys.\ Rev.\ Lett.}\ }\textbf
  {\bibinfo {volume} {118}},\ \bibinfo {pages} {048002} (\bibinfo {year}
  {2017})}\BibitemShut {NoStop}%
\bibitem [{\citenamefont {Germs}\ \emph {et~al.}(2013)\citenamefont {Germs},
  \citenamefont {Roeling}, \citenamefont {van IJzendoorn}, \citenamefont
  {Janssen},\ and\ \citenamefont {Kemerink}}]{germs13}%
  \BibitemOpen
  \bibfield  {author} {\bibinfo {author} {\bibfnamefont {W.~C.}\ \bibnamefont
  {Germs}}, \bibinfo {author} {\bibfnamefont {E.~M.}\ \bibnamefont {Roeling}},
  \bibinfo {author} {\bibfnamefont {L.~J.}\ \bibnamefont {van IJzendoorn}},
  \bibinfo {author} {\bibfnamefont {R.~A.~J.}\ \bibnamefont {Janssen}}, \ and\
  \bibinfo {author} {\bibfnamefont {M.}~\bibnamefont {Kemerink}},\ }\href@noop
  {} {\bibfield  {journal} {\bibinfo  {journal} {Appl.\ Phys.\ Lett.}\ }\textbf
  {\bibinfo {volume} {102}},\ \bibinfo {pages} {073104} (\bibinfo {year}
  {2013})}\BibitemShut {NoStop}%
\bibitem [{\citenamefont {Shinagawa}\ and\ \citenamefont
  {Sasaki}(2016)}]{shinagawa16}%
  \BibitemOpen
  \bibfield  {author} {\bibinfo {author} {\bibfnamefont {R.}~\bibnamefont
  {Shinagawa}}\ and\ \bibinfo {author} {\bibfnamefont {K.}~\bibnamefont
  {Sasaki}},\ }\href@noop {} {\bibfield  {journal} {\bibinfo  {journal} {J.\
  Phys.\ Soc.\ Jpn.}\ }\textbf {\bibinfo {volume} {85}},\ \bibinfo {pages}
  {064004} (\bibinfo {year} {2016})}\BibitemShut {NoStop}%
\bibitem [{\citenamefont {J{\"u}licher}\ \emph {et~al.}(1997)\citenamefont
  {J{\"u}licher}, \citenamefont {Ajdari},\ and\ \citenamefont
  {Prost}}]{julicher97}%
  \BibitemOpen
  \bibfield  {author} {\bibinfo {author} {\bibfnamefont {F.}~\bibnamefont
  {J{\"u}licher}}, \bibinfo {author} {\bibfnamefont {A.}~\bibnamefont
  {Ajdari}}, \ and\ \bibinfo {author} {\bibfnamefont {J.}~\bibnamefont
  {Prost}},\ }\href@noop {} {\bibfield  {journal} {\bibinfo  {journal} {Rev.\
  Mod.\ Phys.}\ }\textbf {\bibinfo {volume} {69}},\ \bibinfo {pages} {1269}
  (\bibinfo {year} {1997})}\BibitemShut {NoStop}%
\bibitem [{\citenamefont {Reimann}(2002)}]{reimann02pr}%
  \BibitemOpen
  \bibfield  {author} {\bibinfo {author} {\bibfnamefont {P.}~\bibnamefont
  {Reimann}},\ }\href@noop {} {\bibfield  {journal} {\bibinfo  {journal}
  {Phys.\ Rep.}\ }\textbf {\bibinfo {volume} {361}},\ \bibinfo {pages} {57}
  (\bibinfo {year} {2002})}\BibitemShut {NoStop}%
\bibitem [{\citenamefont {Kawaguchi}\ \emph {et~al.}(2014)\citenamefont
  {Kawaguchi}, \citenamefont {Sasa},\ and\ \citenamefont
  {Sagawa}}]{kawaguchi14}%
  \BibitemOpen
  \bibfield  {author} {\bibinfo {author} {\bibfnamefont {K.}~\bibnamefont
  {Kawaguchi}}, \bibinfo {author} {\bibfnamefont {S.-I.}\ \bibnamefont {Sasa}},
  \ and\ \bibinfo {author} {\bibfnamefont {T.}~\bibnamefont {Sagawa}},\
  }\href@noop {} {\bibfield  {journal} {\bibinfo  {journal} {Biophys.\ J.}\
  }\textbf {\bibinfo {volume} {106}},\ \bibinfo {pages} {2450} (\bibinfo {year}
  {2014})}\BibitemShut {NoStop}%
\bibitem [{\citenamefont {Harms}\ and\ \citenamefont
  {Lipowsky}(1997)}]{harms97}%
  \BibitemOpen
  \bibfield  {author} {\bibinfo {author} {\bibfnamefont {T.}~\bibnamefont
  {Harms}}\ and\ \bibinfo {author} {\bibfnamefont {R.}~\bibnamefont
  {Lipowsky}},\ }\href@noop {} {\bibfield  {journal} {\bibinfo  {journal}
  {Phys.\ Rev.\ Lett.}\ }\textbf {\bibinfo {volume} {79}},\ \bibinfo {pages}
  {2895} (\bibinfo {year} {1997})}\BibitemShut {NoStop}%
\bibitem [{\citenamefont {Sasaki}(2004)}]{sasaki03}%
  \BibitemOpen
  \bibfield  {author} {\bibinfo {author} {\bibfnamefont {K.}~\bibnamefont
  {Sasaki}},\ }\href@noop {} {\bibfield  {journal} {\bibinfo  {journal} {J.\
  Phys.\ Soc.\ Jpn.}\ }\textbf {\bibinfo {volume} {72}},\ \bibinfo {pages}
  {2497} (\bibinfo {year} {2004})}\BibitemShut {NoStop}%
\bibitem [{\citenamefont {Howard}(2001)}]{howard01}%
  \BibitemOpen
  \bibfield  {author} {\bibinfo {author} {\bibfnamefont {J.}~\bibnamefont
  {Howard}},\ }\href@noop {} {\emph {\bibinfo {title} {Mechanism of Motor
  Proteins and the Cytoskeleton}}}\ (\bibinfo  {publisher} {Sinauer
  Associates},\ \bibinfo {year} {2001})\BibitemShut {NoStop}%
\bibitem [{\citenamefont {Svoboda}\ \emph {et~al.}(1994)\citenamefont
  {Svoboda}, \citenamefont {Mitra},\ and\ \citenamefont {Block}}]{svoboda94}%
  \BibitemOpen
  \bibfield  {author} {\bibinfo {author} {\bibfnamefont {K.}~\bibnamefont
  {Svoboda}}, \bibinfo {author} {\bibfnamefont {P.~P.}\ \bibnamefont {Mitra}},
  \ and\ \bibinfo {author} {\bibfnamefont {S.~M.}\ \bibnamefont {Block}},\
  }\href@noop {} {\bibfield  {journal} {\bibinfo  {journal} {Proc.\ Natl.\
  Acad.\ Sci.\ U.S.A.}\ }\textbf {\bibinfo {volume} {91}},\ \bibinfo {pages}
  {11782 } (\bibinfo {year} {1994})}\BibitemShut {NoStop}%
\bibitem [{\citenamefont {Schnitzer}\ and\ \citenamefont
  {Block}(1995)}]{schnitzer95}%
  \BibitemOpen
  \bibfield  {author} {\bibinfo {author} {\bibfnamefont {M.~J.}\ \bibnamefont
  {Schnitzer}}\ and\ \bibinfo {author} {\bibfnamefont {S.~M.}\ \bibnamefont
  {Block}},\ }\href@noop {} {\bibfield  {journal} {\bibinfo  {journal} {Cold
  Spring Harbor Symp. Quant. Bilol.}\ }\textbf {\bibinfo {volume} {60}},\
  \bibinfo {pages} {793 } (\bibinfo {year} {1995})}\BibitemShut {NoStop}%
\bibitem [{\citenamefont {van Kampen}(2007)}]{vankampen07}%
  \BibitemOpen
  \bibfield  {author} {\bibinfo {author} {\bibfnamefont {N.~G.}\ \bibnamefont
  {van Kampen}},\ }\href@noop {} {\emph {\bibinfo {title} {Stochastic Processes
  in Physics and Chemistry}}},\ \bibinfo {edition} {3rd}\ ed.\ (\bibinfo
  {publisher} {Elsevier},\ \bibinfo {year} {2007})\ Chap.~\bibinfo {chapter}
  {VI}\BibitemShut {NoStop}%
\bibitem [{\citenamefont {Hu}\ \emph {et~al.}(2010)\citenamefont {Hu},
  \citenamefont {Cheng},\ and\ \citenamefont {Berne}}]{hu10}%
  \BibitemOpen
  \bibfield  {author} {\bibinfo {author} {\bibfnamefont {Z.}~\bibnamefont
  {Hu}}, \bibinfo {author} {\bibfnamefont {L.}~\bibnamefont {Cheng}}, \ and\
  \bibinfo {author} {\bibfnamefont {B.~J.}\ \bibnamefont {Berne}},\ }\href@noop
  {} {\bibfield  {journal} {\bibinfo  {journal} {J.\ Chem.\ Phys.}\ }\textbf
  {\bibinfo {volume} {133}},\ \bibinfo {pages} {034105} (\bibinfo {year}
  {2010})}\BibitemShut {NoStop}%
\bibitem [{\citenamefont {H\"anggi}\ \emph {et~al.}(1990)\citenamefont
  {H\"anggi}, \citenamefont {Talkner},\ and\ \citenamefont
  {Borkovec}}]{hanggi90}%
  \BibitemOpen
  \bibfield  {author} {\bibinfo {author} {\bibfnamefont {P.}~\bibnamefont
  {H\"anggi}}, \bibinfo {author} {\bibfnamefont {P.}~\bibnamefont {Talkner}}, \
  and\ \bibinfo {author} {\bibfnamefont {M.}~\bibnamefont {Borkovec}},\
  }\href@noop {} {\bibfield  {journal} {\bibinfo  {journal} {Rev.\ Mod.\
  Phys.}\ }\textbf {\bibinfo {volume} {62}},\ \bibinfo {pages} {251} (\bibinfo
  {year} {1990})}\BibitemShut {NoStop}%
\end{thebibliography}%
\end{document}